\documentclass[aps,pra,twocolumn,superscriptaddress]{revtex4-1}

\usepackage[british]{babel}
\usepackage{epsfig,amsmath}
\usepackage{graphicx}
\usepackage{dcolumn}
\usepackage[none]{hyphenat}
\usepackage{stmaryrd}
\usepackage{mathrsfs}
\usepackage{pifont}
\usepackage{amsthm}
\usepackage{amssymb}
\usepackage{bm}
\usepackage{latexsym}
\usepackage[colorlinks=true,linkcolor=blue,citecolor=blue]{hyperref}
\usepackage{color}
\usepackage{braket}
\usepackage{tikz}
\usetikzlibrary{quantikz}
\usepackage{subcaption}
\begin{document}
\sloppy

\title{Experimental implementation of Leakage Elimination Operators and subspaces protection}
\author{Beatriz García Markaida}
\affiliation{Department of Theoretical Physics and History of Science, The Basque Country University (UPV/EHU), P.O. Box 644, 48080 Bilbao, Spain}
\author{Lian-Ao Wu}
\affiliation{Department of Theoretical Physics and History of Science, The Basque Country University (UPV/EHU), P.O. Box 644, 48080 Bilbao, Spain}
\affiliation{Ikerbasque, Basque Foundation for Science, 48011 Bilbao, Spain}

\begin{abstract}
     Decoherence-induced leakage errors can potentially damage physical or logical qubits by coupling them to other system levels. Here we report the first experimental implementation of Leakage Elimination Operators (LEOs) that aims to reduce this undermining, and that can be applied alongside universal quantum computing. Using IBM's cloud quantum computer, we have studied three potentially applicable examples of subspaces in two- and three-qubit Hilbert spaces and found that the LEOs significantly suppress leakage.
\end{abstract}

\maketitle
\section{Introduction}
Quantum computation relies on qubits, the fundamental quantum information units. For robustness and error correction purposes, it is commonplace to encode the available physical qubits into \textit{logical qubits}, forming the logical qubit subspace. Ideally, no information would be lost from this subspace into the rest of the qubit system or the environment; however, whenever we deal with a real quantum system, we will encounter non-unitary system dynamical processes arising from system-environment coupling, referred to as \textit{decoherence}. Such decoherence results in a loss of information from the system to the environment, as well as between subspaces of the system. This mixing between subspaces is called \textit{leakage}, and it is a source of severe errors in quantum computing. Decoherence contributes to the decay of quantum coherence and entanglement, which are crucial  in quantum computing and control \cite{Lidar2003}, and, particularly, leakage errors destroy the error protection benefit that we would expect from an encoding of the qubits \cite{Byrd2005}. It is, therefore, unsurprising that decoherence is regarded as the biggest challenge nowadays in quantum information, technologies and quantum computation in particular.

\medskip
Significant theoretical and experimental efforts have been made to overcome the hardships that decoherence produces, both through a quantum circuit model point of view and in the realm of adiabatic quantum computing. The equivalence between these approaches has been proven \cite{Mizel2007}. These efforts include quantum error correction codes (QECCs) \cite{Nielsen}, error avoiding and noiseless quantum codes \cite{Zanardi1997B, Zanardi1997, Zanardi1998, Zanardi2000}, usage of decoherence-free subspaces (DFS) \cite{Lidar1998, Lidar2003, Pyshkin2016}, dynamical decoupling \cite{Pokharel2018, Jing2013B}, strategies to improve quantum adiabatic processes \cite{Wang2018,Jing2016}, control of decoherence through randomized white noise fields \cite{Jing2013}, encondings to get rid of decoherence in solid-state quantum devices \cite{Byrd2002}, or proposals to implement and scale ground-state quantum computers \cite{Mizel2001, Mizel2002, Mizel2004}. However, a universal fault-tolerant quantum computer (QC) is still far from reach, due to complications in experimental implementations of the aforementioned logical qubit-encoding and stabilizers. Most research is thus focused on the application of quantum theory to quantum setups and quantum algorithms, which are of interest for building a future large-scale fault-tolerant QC. 

\medskip
Here we consider three potentially applicable examples of subspaces in two- and three-qubit Hilbert spaces and study leakage suppression via \textit{leakage elimination operators} (LEOs) on IBM's cloud quantum computer.  For the first time, we implement three LEOs on the IBM's QC~\cite{Pokharel2018} and show experimentally that the LEOs can significantly suppress leakage, as predicted in the previous theory~\cite{Wu2002, Wu2009, Byrd2005}. While QECCs are not necessarily compatible with all encodings, LEOs can be compatible with universal quantum computing~\cite{Wu2002}, which allows their application alongside any algorithm running in a QC. 


\section{Universal leakage elimination}\label{theory}
\medskip
Efforts to get rid of decoherence begin with the choice of the logical qubit subspace. If one could, somehow, find an available subspace that remains protected against decoherence, a \textit{decoherence-free subspace}, that would constitute the most adequate choice for the logical qubit subspace. However, such a passive idealization is difficult to find, so typically, one looks for active ways to prevent decoherence. In this work, we aim to make the logical qubit subspace \textit{leakage-free}, that is, we try to suppress leakage from the encoded subspace, or \textit{codespace}, $\mathcal{C}$, to other states of the Hilbert space which reside in the orthogonal complement of the codespace, $\mathcal{C}^{\perp}$ \cite{Byrd2005}, such as in continuous variable systems~\cite{CV}. Elimination of the coupling between these subspaces rids the system from the most pressing decoherence-induced error. 
 Further protection from decoherence, if necessary, could be achieved through, for instance, quantum error-correction codes (QECCs). To obtain these {\em leakage-free subspaces}, we will apply leakage-elimination operators, and we will do this through dynamical decoupling sequences.

\medskip
Leakage elimination operators are based on dynamical decoupling controls. These are control pulses used to average away noise in quantum systems, and can be understood as a projection onto a subspace of the space of operators acting on the system Hilbert space $\mathcal{H}_{S}$ \cite{Byrd2005}. More specifically, we will benefit from the application of fast, intense \textit{bang-bang} (BB) pulses. The briefness and strength of these pulses effectively reduce the system-bath interaction Hamiltonian. Consider a general Hamiltonian,

\begin{equation}
    H=H_{S}+H_{B}+H_{SB},
\end{equation}

\noindent
with $H_{S}$ ($H_{B}$) acting exclusively on the system (bath) and $H_{SB}$, the \textit{system-bath interaction Hamiltonian} coupling the system to the bath. If we are to apply the dynamical decoupling method, that means we will be applying control pulses periodically to our system, leaving an interval $\Delta t$ of free evolution between them. Denoting these control pulses as $U_{i}$, and considering free evolution to be negligible while the control pulses are acting on the system, we obtain an expression for an effective unitary evolution for the combined system-bath after $N$ control pulses \cite{Byrd2005},
\begin{equation}\label{evolution1}
    U_{\text{eff}}\approx\prod_{i=0}^{N-1}U_{i}\exp{[-iH\Delta t]}U_{i}^{\dagger}.
\end{equation}
The control sequences should be applied as fast as possible, and the propagator (\ref{evolution1}) will be exact when $N\to\infty$ and $\Delta t\to 0$. The first order $U_{\text{eff}}$ is described by an effective Hamiltonian,

\begin{equation}
    H_{\text{eff}}\approx\frac{1}{N}\sum_{i=0}^{N-1}U_{i}HU_{i}^{\dagger}.
\end{equation}
Ideally, we would be able to completely eliminate the system-bath interaction for $N\to\infty$. The closer our experimental realization is to an ideal BB pulse scenario, the more efficient our LEO will be. This effectiveness depends mainly on three factors: pulse strength, pulse duration, and the time interval between pulses. The higher the pulse strength and the shorter the time scales, the closer the pulse will be to an ideal BB pulse. More precisely, the effectiveness of an LEO depends on the \textit{integral} of these pulse sequences \cite{Jing2015}, which can be of great interest when considering non-ideal LEOs. However, in this work, we will consider our BB sequences to be ideal.

\medskip
To construct an LEO, we arrange the basis vectors of a $N$-level Hilbert space $\mathcal{H}_{N}$ and assume the first two levels, $\ket{0}$ and $\ket{1}$, correspond to our logical qubits. Then, as is stated in \cite{Wu2002}, we can easily classify all system operators into (a) logical operators $E$ acting on the qubit subspace $\mathcal{C}$, (b) operators acting on the orthogonal complement of this subspace, that is, acting on  $\mathcal{C}^{\perp}$, that thus act entirely outside of the logical qubit subspace, and (c) leakage operators $L$, that connect these two orthogonal subspaces, and that take the following forms:

\begin{equation}\label{eq:subspaces}
    \begin{array}{ccc}
        E=\begin{pmatrix}
        B & 0 \\
        0 & 0
        \end{pmatrix}, & 
        E^{\perp}=\begin{pmatrix}
        0 & 0 \\
        0 & C
        \end{pmatrix}, &
        L=\begin{pmatrix}
        0 & D \\
        F & 0
        \end{pmatrix}.
    \end{array}
\end{equation}
Here $B$, $C$, $D$ and $F$ are blocks of dimensions $2\times 2$, $(N-2)\times(N-2)$, $2\times(N-2)$ and $(N-2)\times 2$, respectively. This decomposition is valid both for physical and logical qubits. The general form of an LEO is \cite{Wu2002},

\begin{equation}\label{LEOmatrix}
    R_{L}=e^{i\varphi}\begin{pmatrix}
    -I & 0 \\
    0 & I
    \end{pmatrix},
\end{equation}

\medskip
\noindent
where the identity blocks $I$ have the dimension of the subspace of the encoded qubits and its orthogonal subspace. That is, the dimension of these blocks matches the ones of the blocks in (\ref{eq:subspaces}). $\varphi$ is a global phase. This operator satisfies the following commutation and anti-commutation relations:

\begin{equation}\label{commuting}
    \begin{array}{cc}
         [R_{L},E]=[R_{L},E^{\perp}]=0, & \{R_{L},L\}=0
    \end{array}
\end{equation}
Since the LEO $R_{L}$ commutes with all operators that act on the encoded qubit subspace, we conclude that LEOs can be applied alongside any logical operation and are, thus, compatible with universal quantum computing. After applying a BB parity-kick sequence, we have

\begin{align}\label{LEOlim}
    \notag\lim_{m\to\infty}&\left(e^{-iH_{\text{SB}}t/m}R_{L}^{\dagger}e^{-iH_{\text{SB}}t/m}R_{L}\right)^{m}=\\
    &=\lim_{m\to\infty}\left(e^{-iH_{\text{SB}}t/m}e^{-iR_{L}^{\dagger}H_{\text{SB}}R_{L}t/m}\right)^{m} \\\notag
    &=e^{-iH_{E}t}e^{-iH_{E^{\perp}}t},
\end{align}

\medskip
\noindent
where $H_{\text{SB}}=H_{E}+H_{E^{\perp}}+H_{L}$ is the system-bath interaction. Here $H_{E}$ and $H_{E^{\perp}}$ correspond to the Hamiltonians acting on the qubit subspace and its orthogonal subspace, respectively, and $H_{L}$ is the Hamiltonian of leakage operators. Note that, taking into account the relations in (\ref{commuting}), and knowing that each of the terms of the Hamiltonian will be composed of operators as the ones described in (\ref{eq:subspaces}) the term $e^{-iR_{L}^{\dagger}H_{\text{SB}}R_{L}t/m}$ becomes $e^{-i(H_{E}+H_{E^{\perp}}-H_{L})t/m}$. The term $e^{-iH_{E^{\perp}}t}$ in (\ref{LEOlim}) acts outside the logical qubit subspace and will thus have no effect on the code of interest. The term $e^{-iH_{E}t}$ does act on the logical qubit subspace and can therefore be the source of logical errors. If necessary, these errors may need further treatment, either by QECCs or additional BB pulses \cite{Byrd2005}. We see, however, that the leakage $H_{L}$  has been eliminated.  For practical purposes, we need only consider $m=1$, and equation (\ref{LEOlim}) will hold up to order $t^2$. The condition $t\ll 1/\omega_{c}$ must also be fulfilled, where $\omega_{c}$ is the bath high-frequency cutoff \cite{Wu2002}.

\medskip
A general choice for LEOs is

\begin{equation}\label{generalLEO}
    R_{L}=\exp{(\pm i\pi\hat{n}\cdot\Vec{\sigma}P)},
\end{equation}

\medskip
\noindent
where $\Vec{\sigma}$ is a vector containing all three Pauli matrices (or $X$, $Y$ and $Z$ logical operations), $\hat{n}$ is a real unit vector, and $P$ is a projector operator onto the qubit subspace. When a canonical logical operation is available, however, this projector becomes redundant and we may write an LEO as in \cite{Byrd2005}:

\begin{equation}\label{generalLEOnoP}
    R_{L}=\exp(-i\pi\sigma_{L}),
\end{equation}

\noindent
where $\sigma_{L}$ is any operation fulfilling $\sigma_{L}^{\dagger}=\sigma_{L}$, $\sigma_{L}^{2}=I$, and $\sigma_{L}\ket{\psi}=0$ for any $\ket{\psi}\in\mathcal{C}^{\perp}$.

\section{Methodology}\label{methodology}
We have experimentally tested three LEOs using IBM's cloud quantum computer. The IBM Q project offers access to several quantum devices; in particular, we have worked with the five-qubit IBMX2 server in Yorktown (see \cite{IBMx2} for technical specifications) $-$ as well as the QASM simulator in the same platform, for testing the code before sending it to the server. The native gates in IBMQ are single qubit rotation $R_{\alpha}(\phi)=\exp{(i\phi/2\sigma_{\alpha})}$, with $\alpha=\{x,y,z\}$ and $\sigma_{\alpha}$ being the Pauli matrices. In practice, we have used $X$ ($\sigma_{x}$), $Z$ ($\sigma_{z}$) and $H$ (Hadamard gate) as single-qubit gates and the two-qubit CNOT (controlled-$X$) gate. For each program, the computer makes 1024 shots. The programs have been written in qiskit \cite{Qiskit}, using the notebooks provided by IBM's platform. 

\medskip
For testing the effectiveness of each LEO, we have proceeded as follows: We first initialize the system to be in a quantum state corresponding to the leakage-free subspace of interest. Then we study the effect of the LEO in the system. For that purpose, we apply the corresponding LEO $\tau$ times, and measure. Similarly, to study the free evolution of the same state, we apply $\tau$ identity gates to a system initialized in the same way. We repeat this process for $\tau\in[1,600]$. Each experiment was then repeated inserting an identity gate between LEO pulses, to study the effect of increased free-evolution intervals in the effectiveness of leakage elimination.

\medskip
The first LEO we tested consists of a sequence of $Z$ gates applied to all qubits in the system, which is the first example presented in \cite{Wu2002}. The protected subspace for the two-qubit $(d=2)$ case is $\{\ket{01},\ket{10}\}$ with $R_{L}^{d}=Z_1Z_2$; in the three-qubit $(d=3)$ case, the protected subspace is $\{\ket{001}, \ket{010}, \ket{100}, \ket{111}\}$ and $R_{L}^{d}=Z_1Z_2Z_3$. In these subspaces, the application of a $Z$ operator for each qubit follows the form given in (\ref{generalLEO}) with $\hat{n}=\hat{u}_{z}$; acting as $I$ in the protected subspace and as $-I$ in its orthogonal subspace. 

\begin{figure}[h]
\begin{subfigure}{\linewidth}
    \centering
    \begin{quantikz}
        \lstick{$\ket{0}$} & \gate{H} & \ctrl{1}\slice{1} & \gate{Z} & \qw ... & \gate{Z} & \meter{z} \\
        \lstick{$\ket{0}$} & \gate{X} & \targ{} & \gate{Z} & \qw ... & \gate {Z}\slice{2} & \meter{z}
    \end{quantikz}
    \caption{}
    \label{fig:init1}
\end{subfigure}
\begin{subfigure}{\linewidth}
    \centering
    \begin{quantikz}
        \lstick{$\ket{0}$} & \gate{H} & \qw & \qw & \ctrl{1}\slice{1} & \gate{Z} & \qw ... &\gate{Z}\slice{2} & \meter{z} \\
        \lstick{$\ket{0}$} & \gate{H} & \gate{X} & \ctrl{1} & \targ{} & \gate{Z} & \qw ... &\gate{Z} & \meter{z} \\
        \lstick{$\ket{0}$} & \gate{X} & \qw & \targ{} & \qw & \gate {Z} & \qw ...&\gate{Z} & \meter{z}
    \end{quantikz}
    \caption{}
    \label{fig:init2}
\end{subfigure} 
\caption{Circuits for the initialization and application of a two-qubit (a) and three-quibit (b) LEO, through $Z$ gates. At  step 1, we have successfully initialized the system in the desired state, and step 2 means the application of the LEO has finished.}
\label{fig:init1-2}
\end{figure}
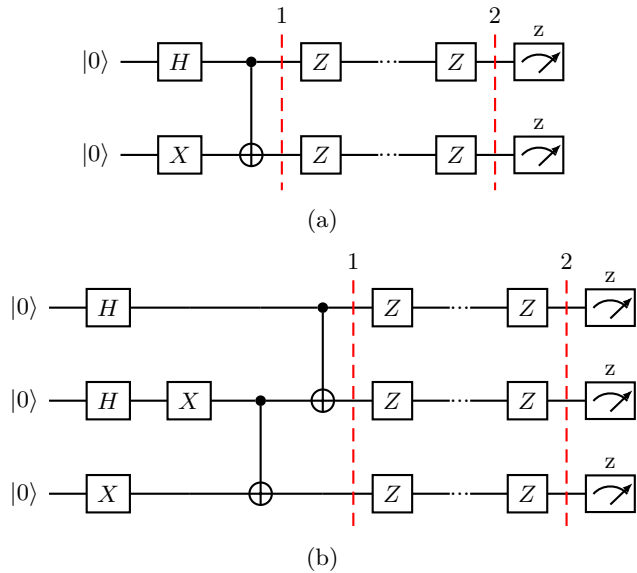

\medskip
For our tests, we have chosen initial states with equal populations in every  state of the protected subspace: $\ket{\Psi_{2}}=(\ket{01}+\ket{10})/\sqrt{2}$ and $\ket{\Psi_{3}}=(\ket{001}+\ket{010}+\ket{100}+\ket{111})/2$, respectively. The gates applied for initialization and LEO application can be seen in the circuits in figures (\ref{fig:init1}) and (\ref{fig:init2}).

\medskip

\begin{figure}[h]
    \centering
    \begin{quantikz}
        \lstick{$\ket{0}$} & \gate{X} & \qw\slice{1} & \ctrl{1} & \qw ... & \ctrl{1}\slice{2} & \qw & \gate{X} & \meter{z} \\
        \lstick{$\ket{0}$} & \gate{H} & \gate{Z} & \targ{} & \qw ... & \targ{} & \gate{H} & \gate{X} &\meter{z}
    \end{quantikz}
    \caption{Circuit for the initialization and application of a two-quibit LEO, through CNOT gates. At  step 1, we have successfully initialized the system in the desired state, and step 2 means the application of the LEO has finished. Note that after the application of the LEO, we must further manipulate our system for measurement purposes.}
    \label{fig:init3}
\end{figure}
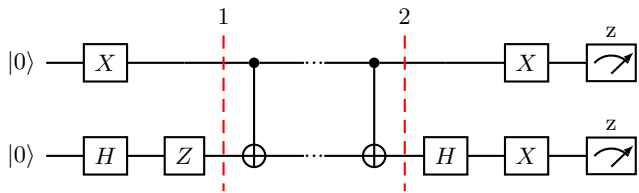

\medskip
The third example considers how a CNOT gate behaves as an LEO.  In this case, the subspace spanned by $\ket{10}-\ket{11}$ is protected from leaking into $\{\ket{00}, \ket{01}, \ket{10}+\ket{11}\}$. The effectiveness of CNOT as an LEO in this system is easy to see if we write its explicit matrix form in the basis of $\{\ket{00}, \ket{01}, \ket{10}+\ket{11},\ket{10}-\ket{11}\}$:

\begin{equation}
    \text{CNOT}=\begin{pmatrix}
    1 & 0 & 0 & 0 \\
    0 & 1 & 0 & 0 \\
    0 & 0 & 1 & 0 \\
    0 & 0 & 0 & -1
    \end{pmatrix}
\end{equation}
This is in accordance with the form of a universal LEO given in (\ref{LEOmatrix}), meaning that CNOT is an LEO in the chosen basis. We prepare the initial state  $\ket{\Phi}=(\ket{10}-\ket{11})/\sqrt{2}$ when studying this LEO. Since the measurement options in IBM's cloud computer are limited to measuring the $z$ component of the spin of the qubits, further manipulation of the system has been needed for measurement purposes $-$ that is, to be able to discern what part of the populations of $\ket{10}$ and $\ket{11}$ correspond to the protected subspace $\{\ket{10}-\ket{11}\}$. After the LEO has been applied $\tau$ times, we have applied a Hadamard gate to the second qubit, and then an $X$ gate to both qubits, so that $\ket{\Phi}$ turns into $\ket{00}$ before the measurement. The circuit describing this can be seen in figure (\ref{fig:init3}).

Once we have completed both runs, we compare the fidelity of the run where an LEO was applied with the one achieved through free evolution. This quantity is obtained for each $\tau$, by computing the ratio of the populations of the desired states by the number of total pulses in each run. For example, in the case of the two-qubit system where we used $Z$ as an LEO, this fidelity would be $ f_{1}={(\text{number of } \ket{01} \text{ states} + \text{number of } \ket{10} \text{ states})}/{1024}$.

\section{Results and discussion}
\begin{figure}[h]
    \begin{subfigure}{0.49\linewidth}
    \centering
    \includegraphics[width=\linewidth]{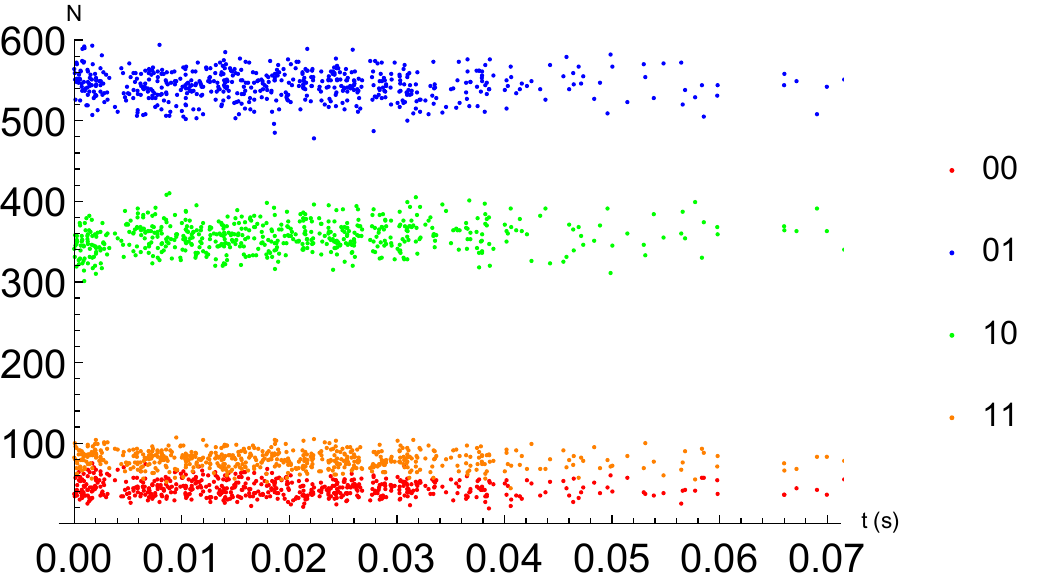}
    \caption{}
    \label{fig:leoZ2}  
    \end{subfigure}
    \begin{subfigure}{0.49\linewidth}
    \centering
    \includegraphics[width=\linewidth]{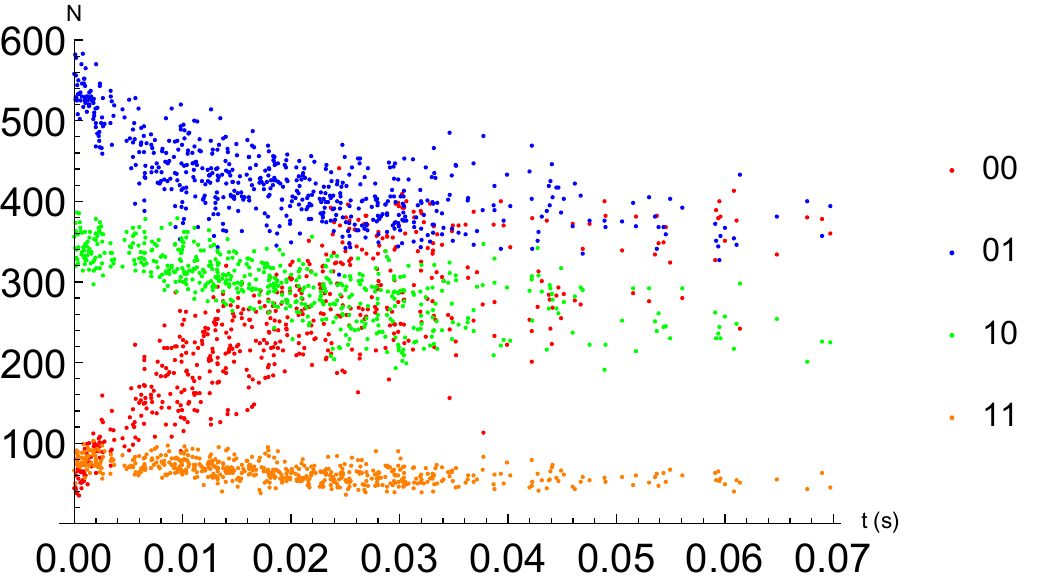}
    \caption{}
    \label{fig:refZ2}
    \end{subfigure}
    \\
    \begin{subfigure}{0.49\linewidth}
    \centering
    \includegraphics[width=\linewidth]{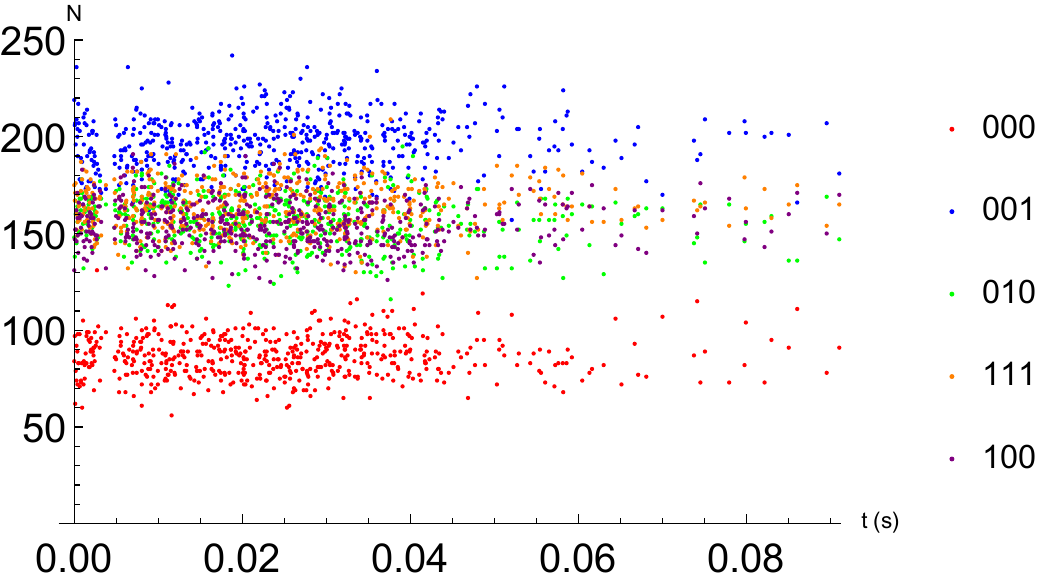}
    \caption{}
    \label{fig:leoZ3}  
    \end{subfigure}
    \medskip
    \begin{subfigure}{0.49\linewidth}
    \centering
    \includegraphics[width=\linewidth]{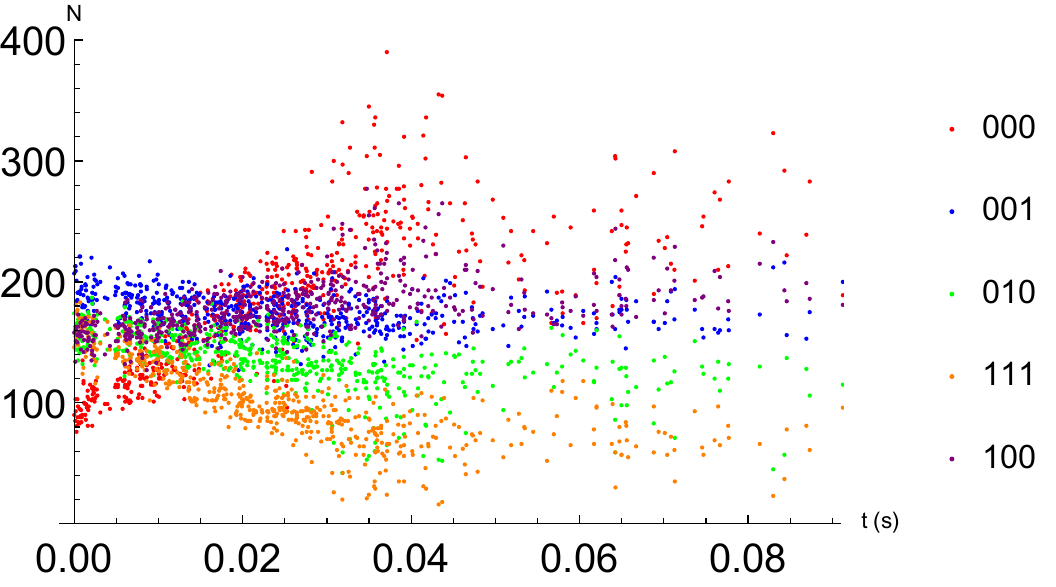}
    \caption{}
    \label{fig:refZ3}
    \end{subfigure}
    \caption{Populations of individual states measured at different times, corresponding to up to 600 pulses, with he application of Z as an LEO  in (a) and (c), and with free evolution in (b) and (d). (a) and (b) show the populations of $\ket{00},\ket{01},\ket{10} \ \text{and} \ \ket{11}$, while (c) and (d) show the populations of $\ket{000}$, $\ket{001}$, $\ket{010}$, $\ket{111}$ and $\ket{100}$ (that is, the populations of the protected subspace states). At each time, 1024 shots where fired.}
    \label{fig:statesZ}
\end{figure}

Figure (\ref{fig:statesZ}) shows the population of each state in the cases where $Z$ was used as an LEO. We can clearly see that, when an LEO is used (see figures (\ref{fig:leoZ2}) and (\ref{fig:leoZ3})), initial populations remain roughly constant over time, with some expected fluctuations due to the probabilistic nature of quantum mechanics. While one would expect populations of the components of the initial state to be evenly distributed (for example, having equal populations of $\ket{01}$ and $\ket{10}$ in (\ref{fig:leoZ2})), differences in the physical qubits prevent this from happening \cite{IBMx2}. When the LEO is not applied (figures (\ref{fig:refZ2}) and (\ref{fig:refZ3})), the population of the ground state increases, as the population of the states of interest decreases. This is due to the decay of qubits in state $\ket{1}$ into the lower energy state $\ket{0}$. This is precisely the leakage we seek to prevent through the use of LEOs.

\begin{figure*}
    \begin{subfigure}{0.45\linewidth}
        \centering
        \includegraphics[width=\linewidth]{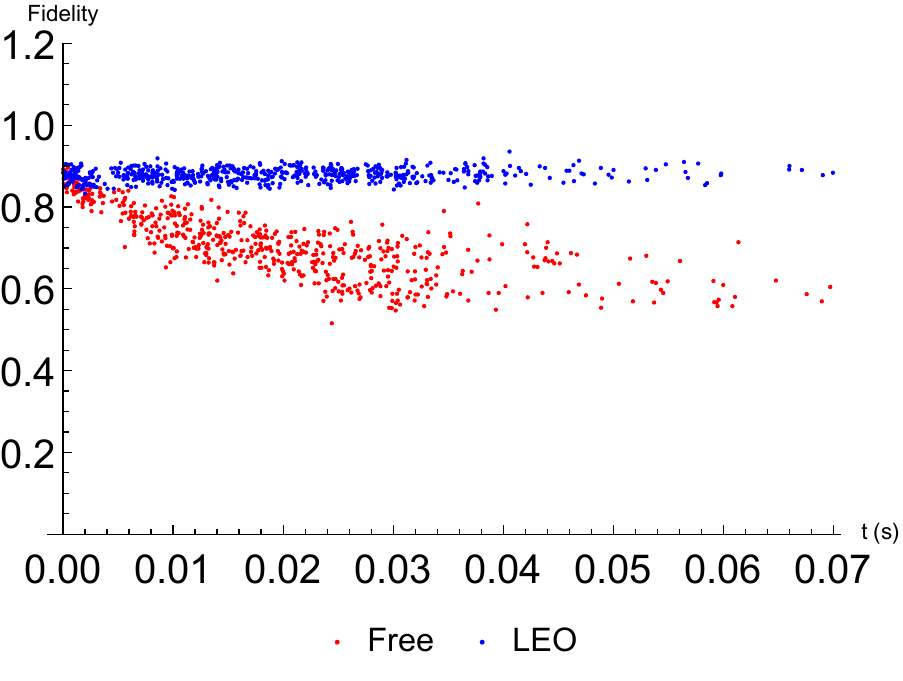}
        \caption{}
        \label{fig:resZ2}  
    \end{subfigure}
    \begin{subfigure}{0.45\linewidth}
        \centering
        \includegraphics[width=\linewidth]{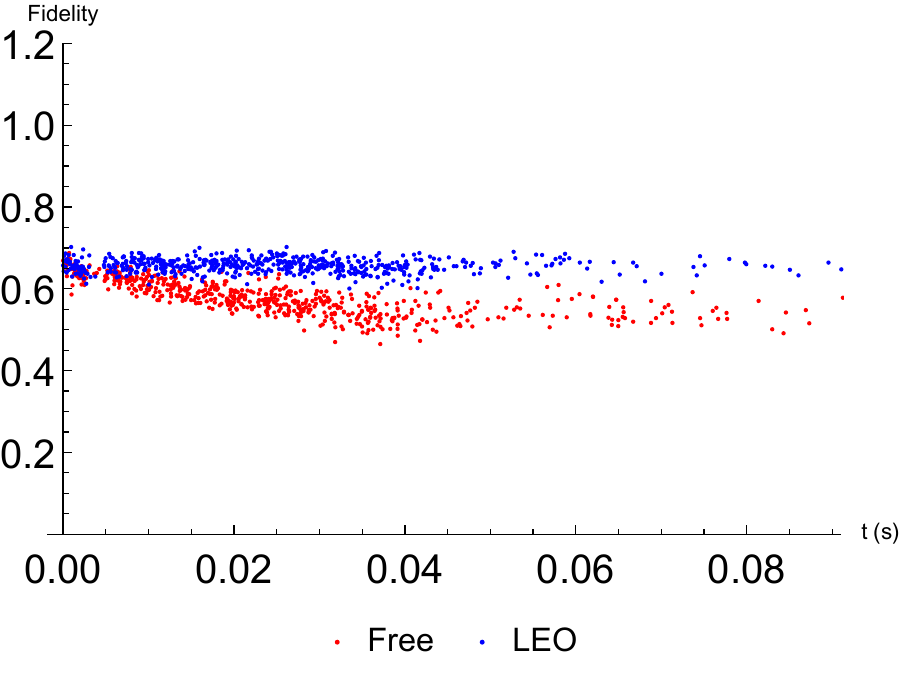}
        \caption{}
        \label{fig:resZ3}
    \end{subfigure}
    \\
    \begin{subfigure}{0.45\linewidth}
        \centering
        \includegraphics[width=\linewidth]{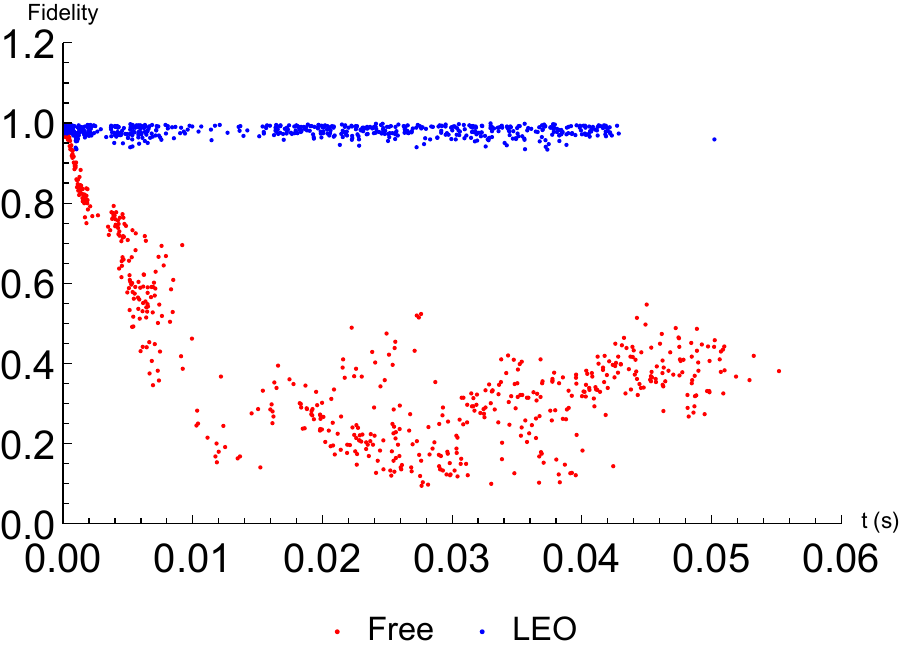}
        \caption{}
        \label{fig:resCNOT}
    \end{subfigure}
    \caption{Fidelity of quantum states measured at different times, corresponding to up to 600 pulses, with 1024 shots at each time. The blue dots correspond to the run where LEO pulses applied; the red dots correspond to the freely evolving system. (a) shows the evolution of a two-qubit system, with inital state $\ket{\Psi_{2}}=(\ket{01}+\ket{10})/\sqrt{2}$ and Z pulses used as an LEO; (b) depicts the three-qubit case that also uses Z as an LEO, with initial state $\ket{\Psi_{3}}=(\ket{001}+\ket{010}+\ket{100}+\ket{111})/2$. In (c) we used $\ket{\Phi}=(\ket{10}-\ket{11})/\sqrt{2}$ as our initial state and the role of LEO was carried out by CNOT gates.}
    \label{fig:res}
\end{figure*}

\medskip
We have found that the performance of the LEOs varies for different examples. It also depends on the dimension of the studied system. All experiments, however, share one trait: application of an LEO maintains the fidelity of the final state constant in time, as we can see in figure (\ref{fig:res}) and predicted in the previous theory~\cite{Wu2002}. This, in accordance to figures (\ref{fig:leoZ2}) and (\ref{fig:leoZ3}), means the population of states in the protected subspace remains constant $-$ that is, \textit{protected}. Depending on the specific LEO at use, we have been able to maintain roughly a $0.8$ or $0.6$ fidelity (see figures (\ref{fig:resZ2}) and (\ref{fig:resZ3})), in the case of using $Z$ gates, or that we may maintain nearly ideal fidelity, in the case of using CNOT gates (figure (\ref{fig:resCNOT})). 

\medskip

\medskip
We see that fidelity at $\tau=1$ is greater for the system where we applied CNOT gates as an LEO; this may be a consequence of the presence of $\ket{00}$ states prior to the manipulation of the final state for measurement purposes. The improvement in fidelity is, nonetheless, the most significant among our three cases of study. Since the time interval between pulses is the same when we use $Z$ as an LEO and when we use CNOT for that purpose (at least in the two-qubit case), the higher efficiency of the latter might also be due to a higher pulse intensity or duration.

\begin{figure*}
    \begin{subfigure}{0.45\linewidth}
        \centering
        \includegraphics[width=0.9\linewidth]{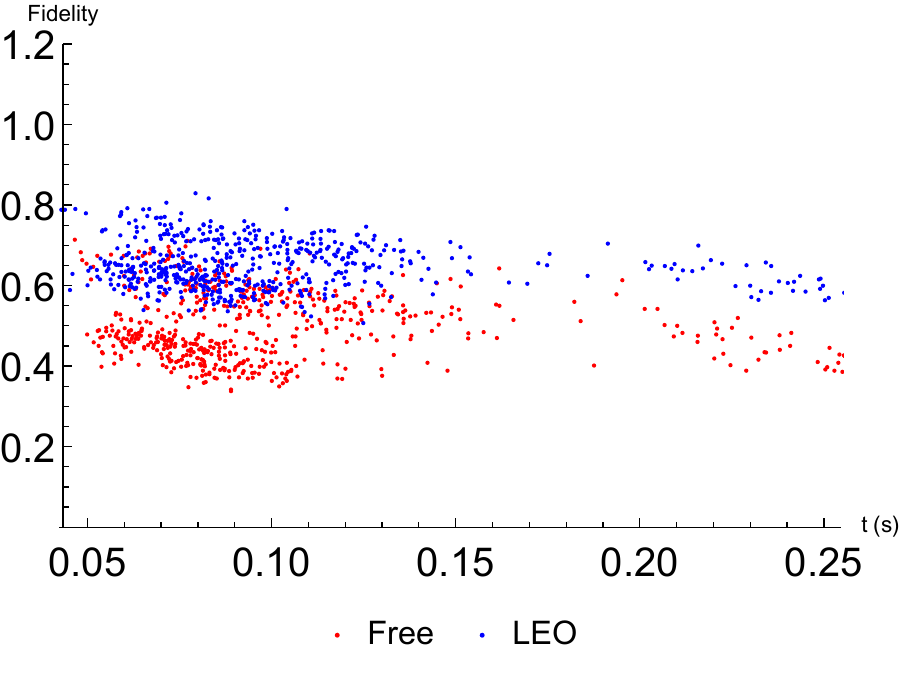}
        \caption{}
        \label{fig:resZID2}  
    \end{subfigure}
    \begin{subfigure}{0.45\linewidth}
        \centering
        \includegraphics[width=0.9\linewidth]{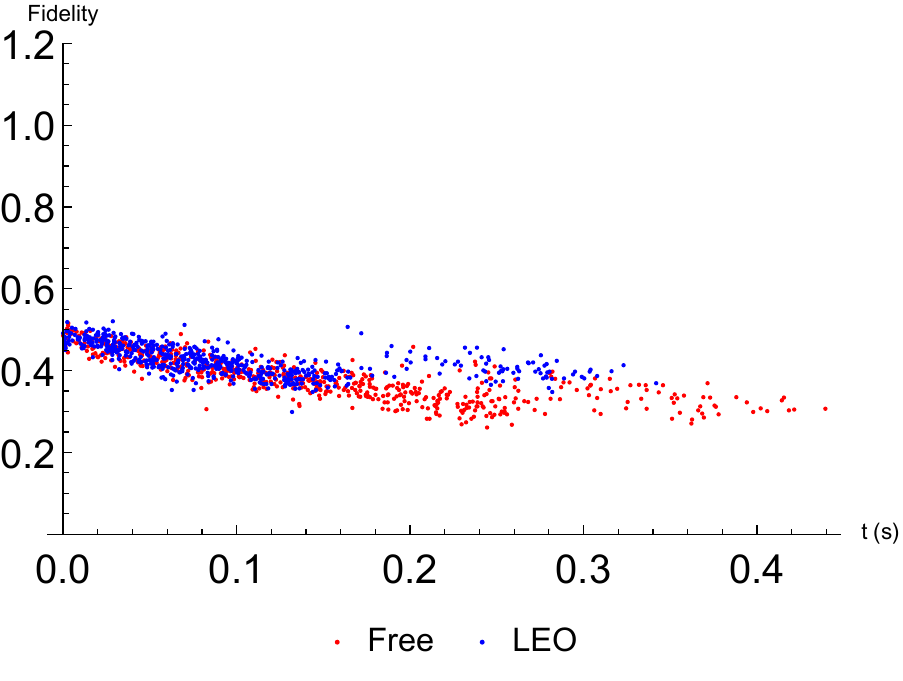}
        \caption{}
        \label{fig:resZID3}  
    \end{subfigure}
    \\
    \begin{subfigure}{0.45\linewidth}
        \centering
        \includegraphics[width=0.9\linewidth]{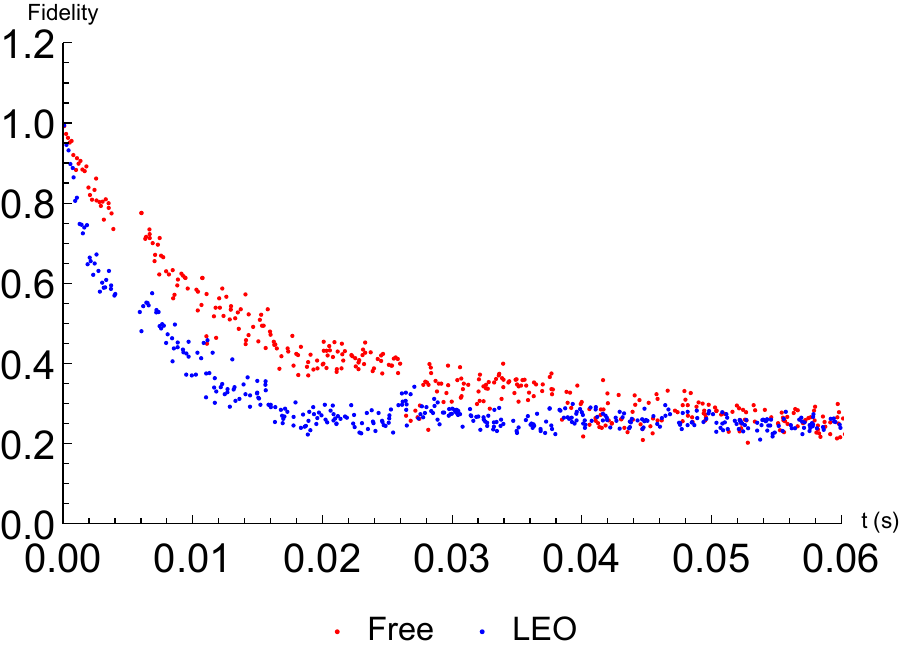}
        \caption{}
        \label{fig:resCNOTID}
    \end{subfigure}
    \caption{Fidelity of quantum states measured at different times, corresponding to up to 600 pulses, with 1024 shots at each time. Blue dots correspond to the run where LEO pulses where applied, with identity pulses between applications; red dots correspond to the freely evolving system. As in figure (\ref{fig:res}), (a) shows the evolution of a two-qubit system, with inital state $\ket{\Psi_{2}}=(\ket{01}+\ket{10})/\sqrt{2}$ and Z pulses used as an LEO; (b) depicts the three-qubit case that also uses Z as an LEO, with initial state $\ket{\Psi_{3}}=(\ket{001}+\ket{010}+\ket{100}+\ket{111})/2$. In (c) we used $\ket{\Phi}=(\ket{10}-\ket{11})/\sqrt{2}$ as our initial state and the role of LEO was carried out by CNOT gates.}
    \label{fig:resID}
\end{figure*}

\medskip
On another note, looking at figure (\ref{fig:resID}), we can see that an increase in free evolution time between LEO pulses decreases the effectiveness of leakage elimination. This is to be expected in theory, since, as explained in section \ref{theory}, the ideal LEO relies on BB pulses for effectiveness. That is, elongating the time between pulses strays us further away from the ideal LEO protocol. This effect is the most prominent for the case of CNOT gates (see figure (\ref{fig:resCNOTID})), which makes the application of an LEO counterproductive.

\medskip

\section{Conclusion}
The present experiments confirm that LEOs can help stabilize fidelity for two- and three-qubit leakage-free subspaces on IBMQ's 5-qubit QC, as predicted theoretically. Although IBMQ's 5-qubit QC limits the application of ideal BB control sequences, further decay in fidelity has been successfully suppressed. We have found that as long as the time interval between pulse applications is short enough (within the capability of IBMQ's 5-qubit QC), the use of an LEO is surely advantageous with respect to a free evolution of the system. Keeping in mind that LEOs are universal and can be applied alongside other qubit operations, we have experimentally given a powerful tool for higher accuracy of quantum algorithms.

\bigskip
This material is based upon work supported by the Basque Country Government (Grant No. IT986-16), and PGC2018-101355- B-I00 (MCIU/AEI/FEDER,UE). We acknowledge use of the IBM Q for this work. The views expressed are those of the authors and do not reflect the official policy or position of IBM or the IBM Q team.

\bibliography{tfm}

\begin{thebibliography}{10}

\bibitem{Lidar2003}
Daniel~A. Lidar and K.~Birgitta Whaley.
\newblock Decoherence-free subspaces and subsystems.
\newblock 2003.
\newblock \href {http://arxiv.org/abs/quant-ph/0301032}
  {\path{arXiv:quant-ph/0301032}}.

\bibitem{Byrd2005}
Mark~S. Byrd, Daniel~A. Lidar, Lian-Ao Wu, and Paolo Zanardi.
\newblock Universal leakage elimination.
\newblock {\em Physical Review A}, 71(5), May 2005.
\newblock \href {https://doi.org/10.1103/physreva.71.052301}
  {\path{doi:10.1103/physreva.71.052301}}.

\bibitem{Mizel2007}
Ari Mizel, Daniel~A. Lidar, and Morgan Mitchell.
\newblock Simple proof of equivalence between adiabatic quantum computation and
  the circuit model.
\newblock {\em Physical Review Letters}, 99(7), August 2007.
\newblock \href {https://doi.org/10.1103/physrevlett.99.070502}
  {\path{doi:10.1103/physrevlett.99.070502}}.

\bibitem{Zanardi1997B}
Paolo Zanardi and Mario Rasetti.
\newblock Error avoiding quantum codes.
\newblock {\em Modern Physics Letters B}, 11(25):1085--1093, October 1997.
\newblock \href {https://doi.org/10.1142/s0217984997001304}
  {\path{doi:10.1142/s0217984997001304}}.

\bibitem{Zanardi1997}
Paolo Zanardi and Mario Rasetti.
\newblock Noiseless quantum codes.
\newblock {\em Physical Review Letters}, 79(17):3306--3309, October 1997.
\newblock \href {https://doi.org/10.1103/physrevlett.79.3306}
  {\path{doi:10.1103/physrevlett.79.3306}}.

\bibitem{Zanardi1998}
Paolo Zanardi.
\newblock Dissipation and decoherence in a quantum register.
\newblock {\em Physical Review A}, 57(5):3276--3284, May 1998.
\newblock \href {https://doi.org/10.1103/physreva.57.3276}
  {\path{doi:10.1103/physreva.57.3276}}.

\bibitem{Zanardi2000}
Paolo Zanardi.
\newblock Stabilizing quantum information.
\newblock {\em Physical Review A}, 63(1), December 2000.
\newblock \href {https://doi.org/10.1103/physreva.63.012301}
  {\path{doi:10.1103/physreva.63.012301}}.

\bibitem{Lidar1998}
Daniel~A. Lidar, I.~L. Chuang, and K.~B. Whaley.
\newblock Decoherence-free subspaces for quantum computation.
\newblock {\em Physical Review Letters}, 81(12):2594--2597, September 1998.
\newblock \href {https://doi.org/10.1103/physrevlett.81.2594}
  {\path{doi:10.1103/physrevlett.81.2594}}.

\bibitem{Pyshkin2016}
P.~V. Pyshkin, Da-Wei Luo, Jun Jing, J.~Q. You, and Lian-Ao Wu.
\newblock Expedited holonomic quantum computation via net zero-energy-cost
  control in decoherence-free subspace.
\newblock {\em Scientific Reports}, 6(1), November 2016.
\newblock \href {https://doi.org/10.1038/srep37781}
  {\path{doi:10.1038/srep37781}}.

\bibitem{Pokharel2018}
Bibek Pokharel, Namit Anand, Benjamin Fortman, and Daniel~A. Lidar.
\newblock Demonstration of fidelity improvement using dynamical decoupling with
  superconducting qubits.
\newblock {\em Physical Review Letters}, 121(22), November 2018.
\newblock \href {https://doi.org/10.1103/physrevlett.121.220502}
  {\path{doi:10.1103/physrevlett.121.220502}}.

\bibitem{Jing2013B}
Jun Jing, Lian-Ao Wu, J.~Q. You, and Ting Yu.
\newblock Nonperturbative quantum dynamical decoupling.
\newblock {\em Physical Review A}, 88(2), August 2013.
\newblock \href {https://doi.org/10.1103/physreva.88.022333}
  {\path{doi:10.1103/physreva.88.022333}}.

\bibitem{Wang2018}
Zhao-Ming Wang, Mark~S. Byrd, Jun Jing, and Lian-Ao Wu.
\newblock Adiabatic leakage elimination operator in an experimental framework.
\newblock {\em Physical Review A}, 97(6), June 2018.
\newblock \href {https://doi.org/10.1103/physreva.97.062312}
  {\path{doi:10.1103/physreva.97.062312}}.

\bibitem{Jing2016}
Jun Jing, Marcelo~S. Sarandy, Daniel~A. Lidar, Da-Wei Luo, and Lian-Ao Wu.
\newblock Eigenstate tracking in open quantum systems.
\newblock {\em Physical Review A}, 94(4), October 2016.
\newblock \href {https://doi.org/10.1103/physreva.94.042131}
  {\path{doi:10.1103/physreva.94.042131}}.

\bibitem{Jing2013}
Jun Jing and Lian-Ao Wu.
\newblock Control of decoherence with no control.
\newblock {\em Scientific Reports}, 3(1), September 2013.
\newblock \href {https://doi.org/10.1038/srep02746}
  {\path{doi:10.1038/srep02746}}.

\bibitem{Byrd2002}
Mark~S. Byrd and Daniel~A. Lidar.
\newblock Comprehensive encoding and decoupling solution to problems of
  decoherence and design in solid-state quantum computing.
\newblock {\em Physical Review Letters}, 89(4), July 2002.
\newblock \href {https://doi.org/10.1103/physrevlett.89.047901}
  {\path{doi:10.1103/physrevlett.89.047901}}.

\bibitem{Mizel2001}
Ari Mizel, M.~W. Mitchell, and Marvin~L. Cohen.
\newblock Energy barrier to decoherence.
\newblock {\em Physical Review A}, 63(4), March 2001.
\newblock \href {https://doi.org/10.1103/physreva.63.040302}
  {\path{doi:10.1103/physreva.63.040302}}.

\bibitem{Mizel2002}
Ari Mizel, M.~W. Mitchell, and Marvin~L. Cohen.
\newblock Scaling considerations in ground-state quantum computation.
\newblock {\em Physical Review A}, 65(2), January 2002.
\newblock \href {https://doi.org/10.1103/physreva.65.022315}
  {\path{doi:10.1103/physreva.65.022315}}.

\bibitem{Mizel2004}
Ari Mizel.
\newblock Mimicking time evolution within a quantum ground state: Ground-state
  quantum computation, cloning, and teleportation.
\newblock {\em Physical Review A}, 70(1), July 2004.
\newblock \href {https://doi.org/10.1103/physreva.70.012304}
  {\path{doi:10.1103/physreva.70.012304}}.

\bibitem{Wu2002}
Lian-Ao Wu, Mark~S. Byrd, and Daniel~A. Lidar.
\newblock Efficient universal leakage elimination for physical and encoded
  qubits.
\newblock {\em Physical Review Letters}, 89(12), August 2002.
\newblock \href {https://doi.org/10.1103/physrevlett.89.127901}
  {\path{doi:10.1103/physrevlett.89.127901}}.

\bibitem{Wu2009}
Lian-Ao Wu, Gershon Kurizki, and Paul Brumer.
\newblock Master equation and control of an open quantum system with leakage.
\newblock {\em Physical Review Letters}, 102(8), February 2009.
\newblock \href {https://doi.org/10.1103/physrevlett.102.080405}
  {\path{doi:10.1103/physrevlett.102.080405}}.

\bibitem{Jing2015}
Jun Jing, Lian-Ao Wu, Mark Byrd, J.{\hspace{0.167em}}Q. You, Ting Yu, and
  Zhao-Ming Wang.
\newblock Nonperturbative leakage elimination operators and control of a
  three-level system.
\newblock {\em Physical Review Letters}, 114(19), May 2015.
\newblock \href {https://doi.org/10.1103/physrevlett.114.190502}
  {\path{doi:10.1103/physrevlett.114.190502}}.

\bibitem{IBMx2}
{5-qubit backend: IBMQX team}, ibm q 5 yorktown backend specification v1.1.0.
\newblock Retrieved from
  \url{https://github.com/Qiskit/ibmq-device-information/tree/master/backends/yorktown/V1}.
\newblock 2019.

\bibitem{Qiskit}
H{\'e}ctor~Abraham et~al.
\newblock Qiskit: An open-source framework for quantum computing, 2019.
\newblock \href {https://doi.org/10.5281/zenodo.2562110}
  {\path{doi:10.5281/zenodo.2562110}}.

\end{thebibliography}
\bibliographystyle{unsrturl}

\end{document}